\documentclass[a4paper]{article}
\usepackage{INTERSPEECH2021}
\usepackage{cite, url}
\usepackage{multicol, multirow}
\usepackage{hyperref}
\usepackage{xcolor}
\usepackage{arydshln}
\usepackage{tabularx}
\usepackage{algorithm}
\usepackage{algpseudocode}
\usepackage[moderate]{savetrees}
\usepackage{footnote}
\usepackage{makecell}
\newcolumntype{Y}{>{\centering\arraybackslash}X}
\hypersetup{
    colorlinks=true,
    linkcolor=purple,
    filecolor=magenta,      
    urlcolor=purple,
}
\newcommand{\newpara}[1]{\vspace{8pt}\noindent\textbf{#1}}

\title{Adapting Speaker Embeddings for Speaker Diarisation}
\name{Youngki Kwon, Jee-weon Jung, Hee-Soo Heo, You Jin Kim, Bong-Jin Lee, Joon Son Chung}
\address{Naver Corporation, South Korea}
\email{youngki.kwon@navercorp.com}
\begin{document}
\maketitle
\begin{abstract}
The goal of this paper is to adapt speaker embeddings for solving the problem of speaker diarisation.
The quality of speaker embeddings is paramount to the performance of speaker diarisation systems. 
Despite this, prior works in the field have directly used embeddings designed only to be effective on the speaker verification task.
In this paper, we propose three techniques that can be used to better adapt the speaker embeddings for diarisation: dimensionality reduction, attention-based embedding aggregation, and non-speech clustering.
A wide range of experiments is performed on various challenging datasets. 
The results demonstrate that all three techniques contribute positively to the performance of the diarisation system achieving an average relative improvement of 25.07\% in terms of diarisation error rate over the baseline.
\end{abstract}

\section{Introduction}
Speaker diarisation addresses the task of segmenting multi-speaker audio recordings into homogeneous single speaker segments, effectively solving \textit{“who spoke when”}. The task is a valuable pre-processing step for understanding and transcribing conversations. The ability to determine who spoke when facilitates a rich transcription of conversation providing context as well as the content.

Recent works on speaker diarisation can be divided into two strands. The first strand trains speaker diarisation systems in an end-to-end manner~\cite{park2021review}.
Specifically, several attempts have been made to directly perform speaker diarisation from audio input by constructing deep neural networks (DNNs) with multiple recurrent layers and training them with permutation invariant training technique~\cite{horiguchi2021hitachi, xue2021online, horiguchi2020end}. 
However, the existing end-to-end systems only show good performance in constrained settings and do not generalise well to real-world conditions. 

The second strand of work utilises the traditional pipeline which typically comprises separate modules for each of the steps that must be performed for speaker diarisation. 
Although the exact specification differs from one work to another, the majority of works use at least three independent components: a speech activity detection (SAD) module that detects speech segments from an input audio, an embedding extraction module that extracts speaker representations ({\em i.e.} embeddings), and a clustering module that maps the extracted embeddings to the clusters of unknown numbers~\cite{garcia2017speaker, han2008strategies}. The weakness of the approach is that the individual components, particularly the SAD module and the speaker embedding extractor, are pre-trained and not optimised for the diarisation task.

Despite this, most of the best performing systems in recent challenges are based either on the traditional pipeline or a hybrid of the two types (traditional and end-to-end systems) \cite{ryant2020third, watanabe2020chime}. 
Moreover, recent state-of-the-art approaches based on the Bayesian hidden Markov model \cite{diez2020optimizing, diez2019bayesian, diez2018speaker} and target speaker voice activity detection \cite{medennikov2020target} also require initial diarisation results to bootstrap their operation, which highlights the importance of the traditional baseline system. 
The focus of this work will therefore be on improving the embeddings for the traditional pipeline.% similar to \cite{wang2020speaker}.

To this end, we propose two additional steps on top of traditional speaker diarisation systems. 
First, we propose techniques that enhance the existing embedding to be suitable for the speaker diarisation task. 
% \youngki{Wang et al.~\cite{wang2020speaker} use graph neural networks to improve speaker embeddings in diarization. Still, in most existing systems, the embeddings trained for speaker verification are directly used for diarisation.} 
In the existing systems, the embeddings trained for speaker verification are directly used for diarisation.
However, there should be differences in the use of features, despite the fact that the embeddings should share similar representations of speakers.
For example, the speaker verification task requires discrimination for more than thousands of speakers at a time, whereas the diarisation task requires discrimination only for ten or fewer speakers.
There is also a difference in that the duration for comparing speakers is shorter in diarisation compared to speaker verification.
Typically, the VoxCeleb datasets \cite{Nagrani19, Chung18b, Nagrani17}, which are widely used in speaker verification, use speech segments over ten seconds on average in order to compare speakers. 
On the other hand, in diarisation, the comparison should be performed at shorter intervals, for example between one or two seconds due to the issue of unknown speaker changing points. 
% \youngki{Considering the differences between these tasks, there are studies to adapt speaker embedding for diarisation task. Wang et al. use graph neural networks (GNN) to improve speaker embedding in diarisation. However, training GNN requires reference RTTM as a label and manual annotation for diarisation requires expensive process.
% Instead, we propose embedding adaptation techniques working in an unsupervised manner. Those techniques reduce the dimension of the existing embedding and enhance the representative power of the speaker within a session, which are described in Sections~\ref{ssec:DR} and \ref{ssec:AA} respectively.}
Considering the differences between these tasks, we propose to reduce the dimensionality of existing embeddings and to enhance the representative power of the speaker within a session, further described in Sections~\ref{ssec:DR} and \ref{ssec:AA} respectively. 

Second, we propose a technique referred to as the non-speech clustering that complements existing SAD.
% As the very first module in speaker diarisation, small changes in the SAD module can greatly affect the overall system performance~\cite{anguera2012speaker}.
In existing works, the SAD module is trained and adopted independently of feature extraction and clustering steps.
This could cause inefficiency in the overall pipeline, considering that the operations of other modules highly depend on the results from SAD performed first.
We train the embedding extractor to represent non-speech segments as well as different speakers, as described in Section~\ref{ssec:NS}.
Not only does this improve the SAD performance by introducing an ensemble-like effect with the existing SAD module, it also helps to reduce speaker confusion error by excluding less confident embeddings in the clustering process.
% In order to handle this problem, a method of combining the SAD module, embedding extraction, and clustering process is proposed and explained in Section~\ref{ssec:NS}.

Experiments are performed on two datasets from the previous DIHARD challenges and an internal dataset of real-world conversations. The proposed methods consistently demonstrate significant improvements over baselines across many different settings, and the performances of our best systems exceed the state-of-the-art in many of the challenge sub-tasks.

\section{Baseline system}
\label{sec:conventional}

This section describes the baseline diarisation system for the experiments in this paper.

\subsection{Speech activity detection}
\label{ssec:SAD}
Speech activity detection (SAD) detects the onsets and offsets of the continuous speech segments in the input audio session.
Our baseline SAD module is identical to \cite{landini2020analysis}. 
First, SAD is performed on 25 millisecond windows of 40-dimensional mel-filterbank features, shifting 10 millisecond at a time.
To utilise short-term context information, we stack five frames from both left and right to form a 440-dimensional input feature.
The extracted features are fed into a multi-layer perceptron (MLP) with 3 hidden layers to decide whether each frame contains speech.
Each hidden layer in the MLP network contains 512 nodes activated by the leaky ReLU function.
The MLP network is trained using the development sets of DIHARD I \cite{ryant2018first}, DIHARD II \cite{Ryant2019TheBaselines} and DIHARD III \cite{ryant2020third}, as well as the VoxConverse \cite{chung2020spot} and AMI \cite{carletta2005ami} datasets.

Based on the SAD results, we split each session into continuous speech segments by compensating for the excessively rapid changes in SAD results following \cite{johnston2012webrtc}. 
We decide the onsets and offsets by sliding a window of a certain size (100ms in our configuration). 
In particular, the onset is identified when the ratio of speech-activated frames exceeds 70\% in the window, and the offset is also identified following the same rule for non-speech frames. 
% Based on empirical results, we fix the sizes of the sliding window to ten (100 ms).

\subsection{Speaker embedding extraction}

In this step, we extract the features representing speaker characteristics from the speech segments detected in the previous step.
For the embedding extraction, we train the {\bf H / ASP} architecture described in~\cite{kwon2020ins} with a few modifications \cite{he2016deep, chung2020defence, he2016identity, heo2020clova}.
First, we use the development set of both VoxCeleb 1 and 2~\cite{Nagrani17,Chung18b}.
The number of filters in the first convolutional layer is configured to 64.
We adopt an average pooling instead of attentive statistics pooling for aggregation.
The angular margin softmax \cite{deng2019arcface} objective function is used to train the model.

To extract a speaker embedding from a segment, we first extract a 256-dimensional speaker embedding using a window of 1.5 second width and 0.5 second shift and then apply average to the extracted embeddings.

\subsection{Clustering}
% In this step, we group the embeddings to generate speaker labels for each speech segment using two well-known clustering algorithms; agglomerative hierarchical clustering (AHC) and spectral clustering \cite{day1984efficient, von2007tutorial}. 
Using the extracted speaker embeddings, we then generate speaker labels for each speech segment using two well-known clustering algorithms; agglomerative hierarchical clustering (AHC) and spectral clustering (SPC) \cite{day1984efficient, von2007tutorial}.
AHC constructs a hierarchy based on the distances between features and forms a group. 
For AHC, it is common to manually set a distance threshold to estimate the number of speakers. 
As an alternative, a silhouette coefficient-based trick can be applied to estimate the number of speakers with the expectation of better generalization performance for various environments \cite{kwon2020look, rousseeuw1987silhouettes}.

SPC groups the features using the manifold of embedding space \cite{von2007tutorial, ning2006spectral}. 
First, the affinity matrix where each element represents cosine similarity between two features is calculated. 
Then, we apply eigen-decomposition to the affinity matrix without any additional refinement process~\cite{Heo2021NAVERCHALLENGE}. 
% To determine the number of clusters, eigen-values greater than 20 are counted. 
A threshold of 20 is empirically applied to the eigen-values for determining the number of clusters.
Finally, we perform k-means clustering on the spectral embeddings, which is a set of eigen-vectors corresponding to the largest eigen-values, to estimate the final cluster labels. 

\section{Proposed speaker embedding adaptation}
\label{sec:adap}
\begin{figure}[!t]
 \centering 
 \includegraphics[width=0.55\linewidth,trim=5 5 5 15, clip]{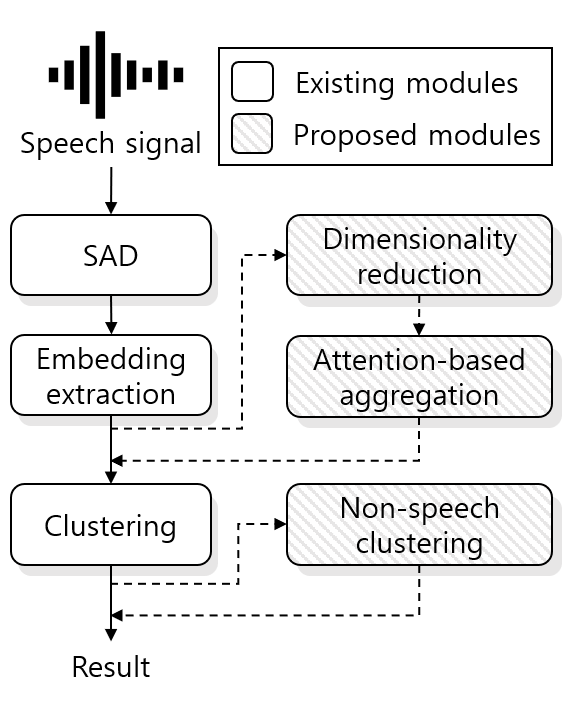}
 \caption{Process pipeline of common speaker diarisation system with the proposed modules. }
 \label{fig:block_diagram}
 \vspace{-2pt}
\end{figure}

% All three techniques are applied sequentially along with the existing steps. 
% This section introduces three proposed techniques for speaker diarisation. 
% Figure \ref{fig:block_diagram} illustrates the entire process pipeline including the proposed techniques. 
% Since each step is designed independently, it would be applied altogether, but only a few necessary steps could be applied depending on the system configuration.
This section introduces three proposed techniques.
Figure \ref{fig:block_diagram} illustrates modified process pipeline including these techniques. 
All three techniques can be independently adopted. 
Thus, certain combination of these techniques can be applied to the existing pipeline. 

\subsection{Dimensionality reduction}
\label{ssec:DR}
% Dimensionality reduction is common in the machine learning field, however, it has not been explored for speaker diarisation to the best of our knowledge. 
% In addition, it common that the speaker embedding extractor, trained for speaker verification task, is applied to speaker diarisation without any modification. 
% Based on our analysis, the embeddings for the speaker diarisation require discriminative power for only a small number of speakers, especially the speakers contained in one session. 
% However, since there are the differences in the required characteristics between two tasks, it would be beneficial to adapt the speaker embeddings for diarisation task.

The majority of studies on speaker diarisation adopt speaker embedding extractors trained for speaker verification without any modifications.
However, required characteristics between two tasks differ, thus, it would be beneficial to adapt the speaker embeddings for diarisation task.

% For speaker diarisation, embeddings require discriminative power only for a small number of speakers contained in a single session.
% This is in contrast to the embeddings for the speaker verification which demands discrimination of thousand of speakers. 
% Therefore, the embeddings extracted for speaker verification may be too high-dimensional and sparse for speaker diarisation. 
In speaker diarisation, embeddings are used to represent only a small number of speakers in one session, different from those in speaker verification required to represent unlimited number of speakers. 
We argue that the embeddings extracted for speaker verification might be excessively high-dimensional, thus, sparse for speaker diarisation. 
For instance, only a small part of the information included in the embeddings would be used to distinguish a small number of speakers, even though the 256-dimensional embeddings are extracted. 
In addition, the remaining information included in the embeddings would have uncertainty due to factors other than the speaker information and cause noise in the affinity matrix. 
To mitigate such a problem, we propose the dimensionality reduction method that finds the suitable low-dimensional code for each session. 
Note that we find the appropriate representation with low dimensionality for each session rather than finding the global representation. 

To achieve this goal, we propose to adopt an auto-encoder (AE) with low-dimensional bottleneck (i.e., code). 
After random initialization, we train the AE to minimise the mean square loss between the original embeddings and the embeddings reconstructed from the network during run-time. 
The AE comprises two layers, one for the encoder and the other for the decoder. 
For the encoder layer, we apply max feature-map activation, well-known for training compact representation~\cite{wu2018light}. 
Using the trained AE, the original embeddings with 256 dimensions are projected into 20-dimensional space. 
We train the AE by 200 epochs using adam optimizer with a 0.001 learning rate for each session.

\subsection{Attention-based embedding aggregation}
\label{ssec:AA}
Despite the success of AHC and SPC in speaker diarisation, these algorithms have inherent limitations depending on the input features.
These limitations should be handled at the embedding-level rather than improving the clustering methods. 
For example, SPC is sensitive to noises in the affinity matrix. 
% AHC is easy to fail to construct hierarchy and estimate the number of clusters wrong affected by outliers in feature space. 
Affected by outliers in embedding space, AHC often fails to construct hierarchy and misestimates the number of clusters.
To mitigate such problems, we propose a feature enhancement technique for speaker diarisation referred to as attention-based embedding aggregation. 
The goal of this technique is to remove noises and outliers that may occur on the affinity matrix using the global context within each session. 

To do so, we aggregate the features of each cluster by using self-attention, which is common in architectures such as Transformer or its variants~\cite{vaswani2017attention,cheng2016long}. 
First, we calculate the attention map for each embedding using the softmax function. 
Subsequently, the embeddings are aggregated based on this attention map. 
These two steps are repeated several times. 
Algorithm \ref{alg:AA} describes the process. 

\begin{algorithm}[!b]
    \caption{Attention-based embedding aggregation}
    \label{alg:AA}
    \begin{algorithmic}[1]
    \State \textbf{Input:} Speaker embeddings \textbf{X} of size \textit{L} $\times$ 20
    \State \textbf{Hyper-parameters:} Number of repetition $N$, Temperature value $\tau$
    \For {$iteration=1,2,\ldots,N$}
			\State Construct affinity matrix \textbf{A}$|$\textbf{A}$_{i,j}$=cos(\textbf{X}$_{i}$,\textbf{X}$_{j}$) 
			\State \textbf{A} = softmax(\textbf{A} *$\tau$) 
			\State \textbf{X} = dot(\textbf{A}, \textbf{X})
	\EndFor
\end{algorithmic}
\end{algorithm}

\begin{figure}[!t]
 \centering 
 \includegraphics[width=1\linewidth,trim=4 4 4 4, clip]{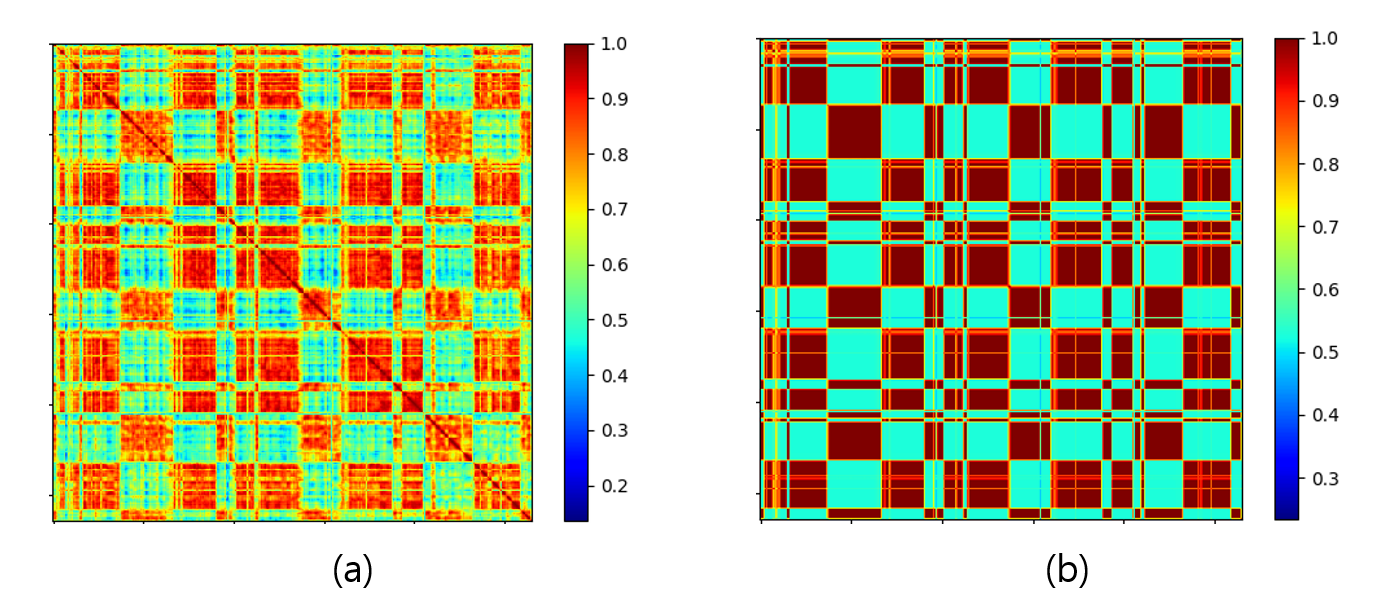}
 \caption{Effect of the attention-based aggregation technique on the affinity matrix. We calculate the affinity matrix using one sample in development set to compare the before (a) and after (b) applying the attention-based aggregation technique. The results show that the proposed technique act like a refinement process that removes the most of noises on the affinity matrix.}
 \label{fig:effect}
 \vspace{-2pt}
 \end{figure}

We expect that the process of aggregating the embeddings for each cluster would remove the noises in the affinity matrix and outliers. 
In the proposed technique, it is necessary to carefully determine the temperature value, which is applied before the softmax function, so that the proper clusters can be formed by aggregation. 
With the appropriate value of temperature and the number of repetitions, we can perform soft version of clustering with this method. 
Figure \ref{fig:effect} compares the effect of the proposed method on the affinity matrix. 
Results show that the proposed method effectively removes the noises on affinity matrix and outliers. 
For more details, two hyper-parameters need to be configured to apply the proposed attention-based aggregation: number of repetitions and temperature value before softmax function. 
We fix these two values as 5 and 15, respectively. 

\subsection{Non-speech clustering}
\label{ssec:NS}
% EPD, which detects speech segments from input audio, is a module that has a great influence on the results of speaker diarisation. 
SAD, the very first building block, has a tremendous impact on speaker diarisation.
In most cases, the existing SAD modules are performed independently from embedding extractor or clustering. 
This design could raise problems, considering the operation of the subsequent steps performed based on the SAD result. 
For example, if the SAD module detects a speech segment where the embedding extractor hardly represents the speaker characteristics, the embedding would have low reliability and degenerates the result of the clustering.

We design the SAD module that closely interacts with the two subsequent steps. 
This approach is motivated from \cite{kwon2020look}.
First, we train the embedding extractor to represent both speech and non-speech segments. 
% To do so, the non-speech segments, which contain only noise without human speech, are assigned to $N+1'th$ where $N$ represents the number of speakers in the training set.
To do so, we add one additional class that includes non-speech regions when training the speaker embedding extractor. 
Here, note that the speaker embedding extractor is trained for speaker identification.
With this process, the embedding extractor learns how to detect not only the specific speaker but also the absence of the speaker.
Therefore, we can handle the non-speech segments as an additional cluster during the clustering step. 

We also propose to combine the results from the SAD module and the clustering output to achieve more reliable results. 
Let the results from the SAD module for each segment be represented as $e_{i}\in{0,1}$ where $0$ and $1$ represent non-speech and speech label, respectively. 
Let $c_{i}\in{0,...,S}$ as the clustering results where $0$ stands the non-speech cluster and $S$ is the number of speakers in one session. 
% We can estimate the reliable segments combining two independent results.
The reliable set $R$ can be constructed if $e_{i}=c_{i}=0$, or $e_{i}=1$ and $c_{i} >0$.
Subsequently, the centroid vector $C_{j}$ for each cluster $j$ is calculated by averaging the embeddings extracted from reliable set $R$. 
Finally, the refined results can be obtained by mapping the embeddings $E_{1,...,T}$ to the nearest centroid among $C_{0,...,S}$. 

The use of more confident embeddings for clustering allows the system to obtain more representative cluster centers, leading to an improvement in speaker confusion error. 
In addition, in case of track 2, we empirically find an ensemble effect of the external SAD module and the embedding-based non-speech clustering contributes to an improvement in FA and MS metrics.

\section{Experiments}
We conduct experiments to evaluate proposed speaker embedding adaptation techniques on three datasets: the first and the second DIHARD challenge datasets \cite{ryant2018first, Ryant2019TheBaselines} and our internal dataset of real-world conversations recorded with a single-channel microphone. 
Sections \ref{ssec:eval_proto} and \ref{ssec:exp_base} describe the evaluation protocol and the baselines, common across all experiments. Section \ref{ssec:exp_dihard_result} describes experiments on each dataset.

\subsection{Evaluation protocol} 
\label{ssec:eval_proto}
We use the diarisation Error Rate (DER) as the primary metric. 
The DER is the sum of three error components: missed speech (MS, a speaker in reference, but not in prediction), false alarm (FA, a speaker in prediction, but not in reference), and speaker confusion error (SC, assigned to wrong speaker ID)
We use the dscore\footnote{https://github.com/nryant/dscore} tool to compute the metrics. 
For the internal dataset, we use a 250ms collar. And for the DIHARD datasets, we use a 0ms collar which is in line with evaluation protocol of the DIHARD challenge.

\subsection{Baselines}
\label{ssec:exp_base}
We test the two baselines described in Section~\ref{sec:conventional}. 
The two baselines each utilise agglomerative hierarchical clustering (AHC) and spectral clustering (SPC). 
Based on the pipelines, we experiment all possible combinations of suggested techniques in Section~\ref{sec:adap}: dimensionality reduction (DR), attention-based embedding aggregation (AA), and non-speech clustering (NS).

% Each configuration requires hyper-parameters, which is set on each dataset. 
%\youngki{The purpose of this section is to show the effect of proposed feature adaptive techniques. Especially, I want to introduce that our method is effective on various pipeline conditions and channels.}
%\youngki{The original plan shows the effect via the experiment result of two DIHARD evaluation sets because the DIHARD dataset is famous and very challenging.}
%\youngki{However, as we mentioned on dihard2 table, our pipeline's score is much lower than the DIHARD2 winner systems. So, we can't use the DIHARD2 result. And we should give a supplementary result to show effectiveness on various channels.}
%\youngki{So, there are three options that we can choose from.}
%\youngki{The first option is using the DIHARD3 development set result. It is jiwon's OSD result. We should explain his OSD in section 2 and compute the score about combinations of dr, aa, and ns. Honestly, I don't know the detail of OSD. So, I need jiwon's help.}
%\youngki{The second option is using voxconverse. It is also hard to beat the winner system of vonconverse challenge. However, if you want the score, I will try it.}
%\youngki{The last option is using an internal dataset. That dataset is already used in our SLT paper 'Look who's not talking.' And I already get the number, and you can check that in table 3. It also shows the same aspect to dihard1's result. I am worried that the result can't persuade a reviewer because the dataset is an 'internal' dataset.}
%\subsection{Result}
%\label{ssec:exp_res}
% 
% \subsection{Datasets} 
\subsection{Results on each dataset} 
\label{ssec:exp_dihard_result}

\begin{table}[t!]
  \centering
  \caption{Results on the DIHARD I challenge dataset using the baseline with every combination of proposed techniques. (FA: false alarm, MS: missed speech, SC: speaker confusion, SPC: Spectral clustering).}
    \resizebox{0.9\linewidth}{!}{\begin{tabular}{lccccc}
    \Xhline{3\arrayrulewidth}
    %\vspace{4pt}
    Configuration  & DER & FA & MS & SC\\
    \midrule
    \multicolumn{5} {c} {\bf Oracle SAD (Track 1)} \\ 
    \midrule
    SPC  & 29.72 & 0.0 & 8.71 & 21.01\\%p0.9, r0.4609
    SPC + DR & 29.16 & 0.0 & 8.71 & 20.45\\%p0.9, r0.4609
    SPC + AA & 19.51 & 0.0 & 8.71 & 10.80\\%p0.9, r0.4609
    SPC + NS & 27.55 & 0.0 & 8.71 & 18.83\\%p0.9, r0.4609
    SPC + DR + AA & 19.12 & 0.0 & 8.71 & 10.41\\%p0.9, r0.4609
    SPC + DR + NS & 20.30 & 0.0 & 8.71 & 11.59\\%p0.9, r0.4609
    SPC + AA + NS & 17.24 & 0.0 & 8.71 & 8.52\\%p0.9, r0.4609
    SPC + DR + AA + NS  & {\bf 16.83} & 0.0 & 8.71 & 8.12\\%p0.9, r0.4609
    \midrule
    AHC  & 21.21 & 0.0 & 8.71 & 12.50\\%p0.9, r0.4609
    AHC + DR & 19.85 & 0.0 & 8.71 & 11.14\\%p0.9, r0.4609
    AHC + AA & 21.42 & 0.0 & 8.71 & 12.71\\%p0.9, r0.4609
    AHC + NS & 20.97 & 0.0 & 8.71 & 12.26\\%p0.9, r0.4609
    AHC + DR + AA & 19.50 & 0.0 & 8.71 & 10.78\\%p0.9, r0.4609
    AHC + DR + NS & 18.45 & 0.0 & 8.71 & 9.74\\%p0.9, r0.4609
    AHC + AA + NS & 21.59 & 0.0 & 8.71 & 12.88\\%p0.9, r0.4609
    AHC + DR + AA + NS  & 17.81 & 0.0 & 8.71 & 9.10\\%p0.9,
    
    \midrule
    Track 1 Winner \cite{sell2018diarization}   & 23.73 & - & - & -\\%p0.9, r0.4609

    \midrule
    \multicolumn{5} {c} {\bf System SAD (Track 2)} \\ 
    \midrule
    SPC   & 49.20 & 16.17 & 10.60 & 22.41\\
    SPC + DR    & 48.74 & 16.17 & 10.60 & 21.96\\%p0.9, r0.4609
    SPC + AA    & 37.01 & 16.17 & 10.60 & 10.22\\%p0.9, r0.4609
    SPC + NS    & 46.15 & 14.55 & 10.26 & 21.33\\%p0.9, r0.4609
    SPC + DR + AA    & 37.43 & 16.17 & 10.60 & 10.65\\%p0.9, r0.4609
    SPC + DR + NS    & 39.67 & 14.52 & 10.37 & 14.77\\%p0.9, r0.4609
    SPC + AA + NS    & {\bf 32.83} & 14.11 & 10.60 & 8.10\\%p0.9, r0.4609
    SPC + DR + AA + NS   & 33.54 & 13.66 & 11.61 & 8.27\\%p0.9, r0.4609
    
    \midrule
    AHC   & 40.24 & 16.17 & 10.60 & 13.45\\
    AHC + DR    & 40.76 & 16.17 & 10.60 & 13.97\\%p0.9, r0.4609
    AHC + AA    & 39.46 & 16.17 & 10.60 & 12.68\\%p0.9, r0.4609
    AHC + NS    & 37.49 & 14.22 & 10.55 & 12.71\\%p0.9, r0.4609
    AHC + DR + AA    & 38.64 & 16.17 & 10.60 & 11.86\\%p0.9, r0.4609
    AHC + DR + NS    & 35.16 & 14.42 & 10.47 & 10.26\\%p0.9, r0.4609
    AHC + AA + NS    & 37.36 & 14.32 & 10.62 & 12.42\\%p0.9, r0.4609
    AHC + DR + AA + NS   & 33.44 & 13.82 & 11.34 & 8.28\\%p0.9, r0.4609
    \midrule
    Track 2 Winner \cite{landini2019but}   & 35.51 & - & - & -\\%p0.9, r0.4609

%     w/ OSD    & 13.93 & 0.62 & 9.00 & 4.31\\%p0.9, r0.4609
%     w/ OSD    & 14.03 & 0.82 & 8.70 & 4.51\\%p0.8, r0.4609
    
%    None               & a & b & c & d \\ 
%    Segment            & 2.24     & 2.20    & 2.14    \\ 
%    Add label           & 2.09    & 1.62    & 1.17    \\ 
%    Segment\&Add label  & \textbf{2.06}     & \textbf{1.52}    & \textbf{1.09}    \\
    \Xhline{3\arrayrulewidth}
  \end{tabular}}
%   \caption{Speaker diarization performance using OSD model. `Segment' refers to applying further segmentation on the EPD segment using OSD and `Add label' refers to allocating second best speaker for OSD segments (FA: false alarm, Conf: confusion).}
  \label{tab:dihard1_abl}
\end{table}
\begin{savenotes}
  \begin{table}[t!]
    \caption{Results on the DIHARD II challenge dataset using the baseline with every combination of proposed techniques. (FA: false alarm, MS: missed speech SC: speaker confusion, SPC: spectral clustering).}
    \centering
    % \begin{tabularx}{\linewidth}{lYYYYY}
    % \resizebox{\columnwidth}{!}{\begin{tabular}{|l|l|l|l|l|}

    \resizebox{0.9\linewidth}{!}{\begin{tabular}{lccccc}
      \Xhline{3\arrayrulewidth}
      %\vspace{4pt}
      Configuration  & DER & FA & MS & SC\\
      \midrule
      \multicolumn{5} {c} {\bf  Oracle SAD (Track 1)} \\ 
      \midrule
      SPC   & 28.43 & 0.0 & 9.68 & 18.74\\%p0.9, r0.4609
      SPC + DR  & 28.61 & 0.0 & 9.68 & 18.93\\%p0.9, r0.4609
      SPC + AA  & 20.53 & 0.0 & 9.68 & 10.85\\%p0.9, r0.4609
      SPC + NS  & 26.77 & 0.0 & 9.68 & 17.08\\%p0.9, r0.4609
      SPC + DR + AA  & 19.70 & 0.0 & 9.68 & 10.02\\%p0.9, r0.4609
      SPC + DR + NS  & 21.08 & 0.0 & 9.68 & 11.39\\%p0.9, r0.4609
      SPC + AA + NS  & 18.84 & 0.0 & 9.68 & 9.15\\%p0.9, r0.4609
      SPC + DR + AA + NS   & {\bf 17.98} & 0.0 & 9.68 & 8.29\\%p0.9, r0.4609
      \midrule
      AHC   & 22.88 & 0.0 & 9.68 & 13.19\\%p0.9, r0.4609
      AHC + DR & 20.68 & 0.0 & 9.68 & 10.99\\%p0.9, r0.4609
      AHC + AA & 23.03 & 0.0 & 9.68 & 13.34\\%p0.9, r0.4609
      AHC + NS & 22.56 & 0.0 & 9.68 & 12.87\\%p0.9, r0.4609
      AHC + DR + AA & 20.10 & 0.0 & 9.68 & 10.41\\%p0.9, r0.4609
      AHC + DR + NS & 18.79 & 0.0 & 9.68 & 9.10\\%p0.9, r0.4609
      AHC + AA + NS & 23.14 & 0.0 & 9.68 & 13.45\\%p0.9, r0.4609
      AHC + DR + AA + NS   & 18.67 & 0.0 & 9.68 & 8.99\\%p0.9, r0.4609
  %    Track 2 baseline \cite{ryant2019second}& 50.12 & - & - & -\\%p0.9, r0.4609
      \midrule
      Track 1 Winner\footnote{http://dihard.ldc.upenn.edu/competitions/73\#results}    & 18.42 & - & - & -\\%p0.9, r0.4609
      
      \midrule
      \multicolumn{5} {c} {\bf System SAD (Track 2)} \\
      \midrule
      SPC   & 51.33 & 19.17 & 11.67 & 20.49\\
      SPC + DR    & 51.17 & 19.17 & 11.67 & 20.33\\%p0.9, r0.4609
      SPC + AA    & 41.10 & 19.17 & 11.67 & 10.26\\%p0.9, r0.4609
      SPC + NS    & 47.15 & 16.02 & 11.76 & 19.36\\%p0.9, r0.4609
      SPC + DR + AA    & 41.19 & 19.17 & 11.67 & 10.35\\%p0.9, r0.4609
      SPC + DR + NS    & 41.50 & 15.88 & 12.00 & 13.61\\%p0.9, r0.4609
      SPC + AA + NS    & 35.78 & 14.98 & 12.32 & 8.47\\%p0.9, r0.4609
      SPC + DR + AA + NS   & {\bf 35.67} & 14.28 & 13.46 & 7.91\\%p0.9, r0.4609
      \midrule
      AHC   & 44.15 & 19.17 & 11.67 & 13.31\\
      AHC + DR    & 44.61 & 19.17 & 11.67 & 13.76\\%p0.9, r0.4609
      AHC + AA    & 43.98 & 19.17 & 11.67 & 13.14\\%p0.9, r0.4609
      AHC + NS    & 40.47 & 15.25 & 11.96 & 13.25\\%p0.9, r0.4609
      AHC + DR + AA    & 42.47 & 19.17 & 11.67 & 11.62\\%p0.9, r0.4609
      AHC + DR + NS    & 37.92 & 15.23 & 12.10 & 10.58\\%p0.9, r0.4609
      AHC + AA + NS    & 40.18 & 15.30 & 12.21 & 12.66\\%p0.9, r0.4609
      AHC + DR + AA + NS   & 35.78 & 14.55 & 13.18 & 8.04\\%p0.9, r0.4609
      
      \midrule
      Track 2 Winner\footnote{http://dihard.ldc.upenn.edu/competitions/74\#results}    & 27.11 & - & - & -\\%p0.9, r0.4609 

  %     w/ OSD    & 13.93 & 0.62 & 9.00 & 4.31\\%p0.9, r0.4609
  %     w/ OSD    & 14.03 & 0.82 & 8.70 & 4.51\\%p0.8, r0.4609
      
  %    None               & a & b & c & d \\ 
  %    Segment            & 2.24     & 2.20    & 2.14    \\ 
  %    Add label           & 2.09    & 1.62    & 1.17    \\ 
  %    Segment\&Add label  & \textbf{2.06}     & \textbf{1.52}    & \textbf{1.09}    \\
      \Xhline{3\arrayrulewidth}
    %   \end{tabularx}
      \end{tabular}}
  %   \caption{Speaker diarization performance using OSD model. `Segment' refers to applying further segmentation on the EPD segment using OSD and `Add label' refers to allocating second best speaker for OSD segments (FA: false alarm, Conf: confusion).}
    \label{tab:dihard2_abl}
  \end{table}
\end{savenotes}

\newpara{DIHARD challenge dataset.}
DIHARD\cite{ryant2018first, Ryant2019TheBaselines} is a series of challenges focusing on `hard' diarisation, where the state-of-the-art systems tend to show poor performance. 
The data includes clinical interviews, child language acquisition recordings, restaurant recordings, and so on. 
We perform experiments on the evaluation sets of the first and the second challenge data.

We compare our pipeline to the winning systems of tracks 1, 2 from the challenge. Track 1 is speaker diarisation beginning from reference speech segment. Track 2 is diarisation from scratch. We analyze the effect of proposed techniques without disturbance by SAD error by comparing our results to the track 1 baseline. 
And we also show the performance of full pipeline including SAD by comparing track 2 results.

The results of each dataset are reported in Tables \ref{tab:dihard1_abl} and \ref{tab:dihard2_abl}, respectively. All of the proposed techniques lead to performance improvement, and the system with all proposed embedding adaptation (+ DR + AA + NS) is our best configuration.
In DIHARD I, our best configuration beats the winning systems in both tracks. 
In DIHARD II, it achieves a better score than the winner of track 1.

\begin{savenotes}
\begin{table}[t!]
%   \tiny
  \vspace{-5pt}
  \centering
  \caption{Results on the internal conversations dataset using the baseline with every combination of proposed techniques. (FA: false alarm, MS: missed speech, SC: speaker confusion).}
  \resizebox{0.9\linewidth}{!}{\begin{tabular}{lccccc}
  \midrule
  Configuration  & DER & FA & MS & SC\\
  \midrule
     AHC   & 24.67 & 4.69 & 4.47 & 15.51\\
  % + DR    & 21.84 & 4.69 & 4.47 & 12.68\\%p0.9, r0.4609
  % + AA    & 25.16 & 4.69 & 4.47 & 16.01\\%p0.9, r0.4609
  % + NS    & 25.54 & 2.41 & 4.13 & 19.01\\%p0.9, r0.4609
  % + DR + AA    & 18.1 & 4.69 & 4.47 & 8.94\\%p0.9, r0.4609
  % + DR + NS    & 21.03 & 2.5 & 4.14 & 14.39\\%p0.9, r0.4609
  % + AA + NS    & 24.09 & 2.44 & 4.09 & 17.56\\%p0.9, r0.4609
  AHC + DR + AA + NS   & {\bf 17.97} & 2.18 & 4.28 & 11.52\\%p0.9, r0.4609
  \midrule
  Baseline from \cite{kwon2020look} & 45.9 & 3.1 & 26.4 & 16.4 \\%p0.9, r0.4609
  Best system from \cite{kwon2020look} & 30.0 & 2.2 & 6.9 & 20.9 \\%p0.9, r0.4609
  
%   w/ OSD    & 13.93 & 0.62 & 9.00 & 4.31\\%p0.9, r0.4609
%   w/ OSD    & 14.03 & 0.82 & 8.70 & 4.51\\%p0.8, r0.4609
  
%  None               & a & b & c & d \\ 
%  Segment            & 2.24     & 2.20    & 2.14    \\ 
%  Add label           & 2.09    & 1.62    & 1.17    \\ 
%  Segment\&Add label  & \textbf{2.06}     & \textbf{1.52}    & \textbf{1.09}    \\
  \midrule
  \end{tabular}}
%   \caption{Speaker diarization performance using OSD model. `Segment' refers to applying further segmentation on the EPD segment using OSD and `Add label' refers to allocating second best speaker for OSD segments (FA: false alarm, Conf: confusion).}
  \label{tab:internal_abl}
  \vspace{-8pt}
\end{table}
\end{savenotes}

\newpara{Internal dataset of real-world conversations.}
This dataset comprises recordings collected across diverse domains such as informal discussions, offline meetings, and Zoom meetings. 
Each conversation is recorded under various conditions, from mobile phones to video conferencing microphone arrays, simulating real-world conditions. The recordings have been labeled professionally by trained annotators. Statistics of this dataset are reported in \cite{kwon2020look}.
Table \ref{tab:internal_abl} gives the result of this dataset. The results are consistent with that of DIHARD I, II.

\section{Conclusions}
% Speaker embeddings have a tremendous impact on the performance in speaker diarisation. 
% However, conventional systems directly use speaker embeddings designed for speaker verification without adaptation towards diarisation. 
% In this paper, we argue that adequate adaptation on speaker embeddings can improve diarisation. 
% To validate our assumption, we introduce three additional techniques designed for adaptation based on our analysis: dimensionality reduction, feature enhancement, and non-speech clustering, and evaluate these techniques using various datasets. 
We propose three techniques to adapt embeddings for solving the diarisation problem based on the hypothesis that conventional approach of directly using speaker verification-oriented embeddings may not be ideal. 
DR reduces the dimensionality of embeddings, AA aggregates embeddings refining the affinity matrix, and NS enables consideration of non-speech regions. 
All three techniques can be independently adopted to the existing process pipeline where applying all three techniques improved the performance the most. 
We find that these proposed techniques demonstrate consistent improvement regardless of both SAD and clustering frameworks.

%\section{Acknowledgements}
%We would like to thank Icksang Han for helpful discussions.
\bibliographystyle{IEEEtran}
\clearpage
\bibliography{shortstrings,mybib}

\begin{thebibliography}{10}
\providecommand{\url}[1]{#1}
\csname url@samestyle\endcsname
\providecommand{\newblock}{\relax}
\providecommand{\bibinfo}[2]{#2}
\providecommand{\BIBentrySTDinterwordspacing}{\spaceskip=0pt\relax}
\providecommand{\BIBentryALTinterwordstretchfactor}{4}
\providecommand{\BIBentryALTinterwordspacing}{\spaceskip=\fontdimen2\font plus
\BIBentryALTinterwordstretchfactor\fontdimen3\font minus
  \fontdimen4\font\relax}
\providecommand{\BIBforeignlanguage}[2]{{%
\expandafter\ifx\csname l@#1\endcsname\relax
\typeout{** WARNING: IEEEtran.bst: No hyphenation pattern has been}%
\typeout{** loaded for the language `#1'. Using the pattern for}%
\typeout{** the default language instead.}%
\else
\language=\csname l@#1\endcsname
\fi
#2}}
\providecommand{\BIBdecl}{\relax}
\BIBdecl

\bibitem{park2021review}
T.~J. Park, N.~Kanda, D.~Dimitriadis, K.~J. Han, S.~Watanabe, and S.~Narayanan,
  ``A review of speaker diarization: Recent advances with deep learning,''
  \emph{arXiv preprint arXiv:2101.09624}, 2021.

\bibitem{horiguchi2021hitachi}
S.~Horiguchi, N.~Yalta, P.~Garcia, Y.~Takashima, Y.~Xue, D.~Raj, Z.~Huang,
  Y.~Fujita, S.~Watanabe, and S.~Khudanpur, ``The hitachi-jhu dihard iii
  system: Competitive end-to-end neural diarization and x-vector clustering
  systems combined by dover-lap,'' \emph{arXiv preprint arXiv:2102.01363},
  2021.

\bibitem{xue2021online}
Y.~Xue, S.~Horiguchi, Y.~Fujita, Y.~Takashima, S.~Watanabe, P.~Garcia, and
  K.~Nagamatsu, ``Online end-to-end neural diarization handling overlapping
  speech and flexible numbers of speakers,'' \emph{arXiv preprint
  arXiv:2101.08473}, 2021.

\bibitem{horiguchi2020end}
S.~Horiguchi, P.~Garcia, Y.~Fujita, S.~Watanabe, and K.~Nagamatsu, ``End-to-end
  speaker diarization as post-processing,'' \emph{arXiv preprint
  arXiv:2012.10055}, 2020.

\bibitem{garcia2017speaker}
D.~Garcia-Romero, D.~Snyder, G.~Sell, D.~Povey, and A.~McCree, ``Speaker
  diarization using deep neural network embeddings,'' in \emph{Proc.
  ICASSP}.\hskip 1em plus 0.5em minus 0.4em\relax IEEE, 2017, pp. 4930--4934.

\bibitem{han2008strategies}
K.~J. Han, S.~Kim, and S.~S. Narayanan, ``Strategies to improve the robustness
  of agglomerative hierarchical clustering under data source variation for
  speaker diarization,'' \emph{IEEE Transactions on Audio, Speech, and Language
  Processing}, vol.~16, no.~8, pp. 1590--1601, 2008.

\bibitem{ryant2020third}
N.~Ryant, P.~Singh, V.~Krishnamohan, R.~Varma, K.~Church, C.~Cieri, J.~Du,
  S.~Ganapathy, and M.~Liberman, ``The third dihard diarization challenge,''
  \emph{arXiv preprint arXiv:2012.01477}, 2020.

\bibitem{watanabe2020chime}
S.~Watanabe, M.~Mandel, J.~Barker, and E.~Vincent, ``Chime-6 challenge:
  Tackling multispeaker speech recognition for unsegmented recordings,''
  \emph{arXiv preprint arXiv:2004.09249}, 2020.

\bibitem{diez2020optimizing}
M.~Diez, L.~Burget, F.~Landini, S.~Wang, and H.~{\v{C}}ernock{\`y},
  ``Optimizing bayesian hmm based x-vector clustering for the second dihard
  speech diarization challenge,'' in \emph{Proc. ICASSP}.\hskip 1em plus 0.5em
  minus 0.4em\relax IEEE, 2020, pp. 6519--6523.

\bibitem{diez2019bayesian}
M.~Diez, L.~Burget, S.~Wang, J.~Rohdin, and J.~Cernock{\`y}, ``Bayesian hmm
  based x-vector clustering for speaker diarization.'' in \emph{Proc.
  Interspeech}, 2019, pp. 346--350.

\bibitem{diez2018speaker}
M.~Diez, L.~Burget, and P.~Matejka, ``Speaker diarization based on bayesian hmm
  with eigenvoice priors.'' in \emph{Proc. Odyssey}, 2018, pp. 147--154.

\bibitem{medennikov2020target}
I.~Medennikov, M.~Korenevsky, T.~Prisyach, Y.~Khokhlov, M.~Korenevskaya,
  I.~Sorokin, T.~Timofeeva, A.~Mitrofanov, A.~Andrusenko, I.~Podluzhny
  \emph{et~al.}, ``Target-speaker voice activity detection: a novel approach
  for multi-speaker diarization in a dinner party scenario,'' in \emph{Proc.
  Interspeech}, 2020.

\bibitem{Nagrani19}
A.~Nagrani, J.~S. Chung, W.~Xie, and A.~Zisserman, ``Voxceleb: Large-scale
  speaker verification in the wild,'' \emph{Computer Science and Language},
  2019.

\bibitem{Chung18b}
J.~S. Chung, A.~Nagrani, and A.~Zisserman, ``Voxceleb2: Deep speaker
  recognition,'' in \emph{Proc. Interspeech}, 2018.

\bibitem{Nagrani17}
A.~Nagrani, J.~S. Chung, and A.~Zisserman, ``Voxceleb: a large-scale speaker
  identification dataset,'' in \emph{Proc. Interspeech}, 2017.

\bibitem{landini2020analysis}
F.~Landini, O.~Glembek, P.~Mat{\v{e}}jka, J.~Rohdin, L.~Burget, M.~Diez, and
  A.~Silnova, ``Analysis of the but diarization system for voxconverse
  challenge,'' \emph{arXiv preprint arXiv:2010.11718}, 2020.

\bibitem{ryant2018first}
N.~Ryant, K.~Church, C.~Cieri, A.~Cristia, J.~Du, S.~Ganapathy, and
  M.~Liberman, ``First dihard challenge evaluation plan,'' \emph{2018, tech.
  Rep.}, 2018.

\bibitem{Ryant2019TheBaselines}
------, ``{The second dihard diarization challenge: Dataset, task, and
  baselines},'' in \emph{Proc. Interspeech}, 2019.

\bibitem{chung2020spot}
J.~S. Chung, J.~Huh, A.~Nagrani, T.~Afouras, and A.~Zisserman, ``Spot the
  conversation: speaker diarisation in the wild,'' in \emph{Proc. Interspeech},
  2020.

\bibitem{carletta2005ami}
J.~Carletta, S.~Ashby, S.~Bourban, M.~Flynn, M.~Guillemot, T.~Hain, J.~Kadlec,
  V.~Karaiskos, W.~Kraaij, M.~Kronenthal \emph{et~al.}, ``The ami meeting
  corpus: A pre-announcement,'' in \emph{International workshop on machine
  learning for multimodal interaction}.\hskip 1em plus 0.5em minus 0.4em\relax
  Springer, 2005, pp. 28--39.

\bibitem{johnston2012webrtc}
A.~B. Johnston and D.~C. Burnett, \emph{WebRTC: APIs and RTCWEB protocols of
  the HTML5 real-time web}.\hskip 1em plus 0.5em minus 0.4em\relax Digital
  Codex LLC, 2012.

\bibitem{kwon2020ins}
Y.~Kwon, H.-S. Heo, B.-J. Lee, and J.~S. Chung, ``The ins and outs of speaker
  recognition: lessons from voxsrc 2020,'' in \emph{Proc. ICASSP}, 2021.

\bibitem{he2016deep}
K.~He, X.~Zhang, S.~Ren, and J.~Sun, ``Deep residual learning for image
  recognition,'' in \emph{Proc. CVPR}, 2016, pp. 770--778.

\bibitem{chung2020defence}
J.~S. Chung, J.~Huh, S.~Mun, M.~Lee, H.~S. Heo, S.~Choe, C.~Ham, S.~Jung, B.-J.
  Lee, and I.~Han, ``In defence of metric learning for speaker recognition,''
  in \emph{Proc. Interspeech}, 2020.

\bibitem{he2016identity}
K.~He, X.~Zhang, S.~Ren, and J.~Sun, ``Identity mappings in deep residual
  networks,'' in \emph{Proc. ECCV}.\hskip 1em plus 0.5em minus 0.4em\relax
  Springer, 2016, pp. 630--645.

\bibitem{heo2020clova}
H.~S. Heo, B.-J. Lee, J.~Huh, and J.~S. Chung, ``Clova baseline system for the
  {VoxCeleb} speaker recognition challenge 2020,'' \emph{arXiv preprint
  arXiv:2009.14153}, 2020.

\bibitem{deng2019arcface}
J.~Deng, J.~Guo, N.~Xue, and S.~Zafeiriou, ``Arcface: Additive angular margin
  loss for deep face recognition,'' in \emph{Proc. CVPR}, 2019, pp. 4690--4699.

\bibitem{day1984efficient}
W.~H. Day and H.~Edelsbrunner, ``Efficient algorithms for agglomerative
  hierarchical clustering methods,'' \emph{Journal of classification}, vol.~1,
  no.~1, pp. 7--24, 1984.

\bibitem{von2007tutorial}
U.~Von~Luxburg, ``A tutorial on spectral clustering,'' \emph{Statistics and
  computing}, vol.~17, no.~4, pp. 395--416, 2007.

\bibitem{kwon2020look}
Y.~Kwon, H.~S. Heo, J.~Huh, B.-J. Lee, and J.~S. Chung, ``Look who's not
  talking,'' in \emph{Proc. SLT}, 2021.

\bibitem{rousseeuw1987silhouettes}
P.~J. Rousseeuw, ``Silhouettes: a graphical aid to the interpretation and
  validation of cluster analysis,'' \emph{Journal of computational and applied
  mathematics}, vol.~20, pp. 53--65, 1987.

\bibitem{ning2006spectral}
H.~Ning, M.~Liu, H.~Tang, and T.~S. Huang, ``A spectral clustering approach to
  speaker diarization,'' in \emph{Ninth International Conference on Spoken
  Language Processing}, 2006.

\bibitem{Heo2021NAVERCHALLENGE}
H.-S. Heo, J.-w. Jung, Y.~Kwon, J.~Kim, J.~Huh, J.~S. Chung, and B.-J. Lee,
  ``{NAVER CLOVA SUBMISSION TO THE THIRD DIHARD CHALLENGE},'' Tech. Rep., 2021.

\bibitem{wu2018light}
X.~Wu, R.~He, Z.~Sun, and T.~Tan, ``A light cnn for deep face representation
  with noisy labels,'' \emph{IEEE Transactions on Information Forensics and
  Security}, vol.~13, no.~11, pp. 2884--2896, 2018.

\bibitem{vaswani2017attention}
A.~Vaswani, N.~Shazeer, N.~Parmar, J.~Uszkoreit, L.~Jones, A.~N. Gomez,
  L.~Kaiser, and I.~Polosukhin, ``Attention is all you need,'' in \emph{NIPS},
  2017.

\bibitem{cheng2016long}
J.~Cheng, L.~Dong, and M.~Lapata, ``Long short-term memory-networks for machine
  reading,'' 2016, pp. 551--561.

\bibitem{sell2018diarization}
G.~Sell, D.~Snyder, A.~McCree, D.~Garcia-Romero, J.~Villalba, M.~Maciejewski,
  V.~Manohar, N.~Dehak, D.~Povey, S.~Watanabe \emph{et~al.}, ``Diarization is
  hard: Some experiences and lessons learned for the jhu team in the inaugural
  dihard challenge.'' in \emph{Proc. Interspeech}, 2018, pp. 2808--2812.

\bibitem{landini2019but}
F.~Landini, S.~Wang, M.~Diez, L.~Burget, P.~Mat{\v{e}}jka,
  K.~{\v{Z}}mol{\'\i}kov{\'a}, L.~Mo{\v{s}}ner, O.~Plchot, O.~Novotn{\`y},
  H.~Zeinali \emph{et~al.}, ``But system description for dihard speech
  diarization challenge 2019,'' \emph{arXiv preprint arXiv:1910.08847}, 2019.

\end{thebibliography}
\end{document}